\documentclass[aps,preprint,epsfig]{revtex4}
\usepackage{graphicx,latexsym,amsmath}
\begin{document}
\title{Harmonically confined, semiflexible polymer in a channel: response to a stretching force and spatial distribution of the endpoints}
\author{Theodore W. Burkhardt}
\affiliation{Department of Physics, Temple University, Philadelphia, PA
19122, USA}

\date{\today}
\begin{abstract}
We consider an inextensible, semiflexible polymer or worm-like chain which is confined in the transverse direction by a parabolic potential and subject to a longitudinal force at the ends, so that the polymer is stretched out and backfolding is negligible. Simple analytic expressions for the partition function, valid in this regime, are obtained for chains of arbitrary length with a variety of boundary conditions at the ends. The spatial distribution of the end points or radial distribution function is also analyzed.
\end{abstract}
\pacs{PACS}
\maketitle

\section{Introduction}
The statistical properties of biological polymers in channels with diameters comparable with the persistence length have been studied in several recent experiments \cite{retal,ksp,kkp,kp,hlgr,chemsocrevpt,chemsocrevlc,wr}. When the channel diameter is much smaller than the persistence length, the polymer is stretched out in the channel, so that its longitudinal length is only slightly shorter than its contour length and there is negligible backfolding.
This is the regime studied theoretically in this paper.

We consider the simplest model for a confined biopolymer -  an inextensible, semiflexible filament or
worm-like chain with persistence length $P$ and contour length $L$, confined in the transverse direction and subject to a longitudinal stretching force or tension $\tau$. If the polymer is tightly confined and/or strongly stretched, the line or filament by which we model it is almost straight. Each such configuration correspond to a single valued
function $\vec{r}(t)$, where $(x,y,t)$ are Cartesian coordinates, and
$\vec{r}=(x,y)$ specifies the transverse displacement of the polymer from the $t$ axis. Since the slope $\vec{v}=d\vec{r}/dt$ satisfies
$|\vec{v}|\ll 1$, the difference $\xi$ between the contour length $L=\int_0^{R_\parallel}dt\left[1+\vec{v}(t)^2\right]^{1/2}$ and the longitudinal length $R_\parallel$ may be replaced by
\begin{equation}
\xi=L-R_\parallel={1\over 2}\int_0^L dt\;\vec{v}(t)^2,\label{extension}
\end{equation}
and the Hamiltonian ${\cal H}$ of the worm-like chain {\cite{explain} simplifies to
\begin{equation}
{{\cal H}\over k_BT}=\int_0^L dt\left[{P\over 2}\left({d^2\vec{r}\over dt^2}\right)^2+{f\over 2}\left({d\vec{r}\over dt}\right)^2+V(\vec{r})\right].\label{Hamiltonian}
\end{equation}
The three terms on the right side of Eq. (\ref{Hamiltonian}) correspond to the bending energy, the potential energy $-\tau(R_\parallel-L)$ associated with the stretching force $\tau=k_BT f$, and the confining potential per unit length, all divided by $k_BT$. For a polymer in a channel with hard walls, $V(\vec{r})$ takes the values $0$ and $\infty$ for $\vec{r}$ inside and outside the channel, respectively.

Note that in Eqs. (\ref{extension}) and (\ref{Hamiltonian}), we have replaced the upper limit $R_\parallel$ in the integrals over $t$ by the contour length $L$, as is correct to leading order in the regime $|\vec{v}|\ll 1$, and in this approximation identify the integration variable $t$ with the arc length. Taking this same point of view, we interpret the path integral $Z=\int Dx\int Dy\thinspace\exp\left(-{\cal H}/k_BT\right)$ with ${\cal H}$ given by Eq. (\ref{Hamiltonian}), studied in detail below, as the partition function of a polymer of fixed contour length $L$ but fluctuating longitudinal length $R_\parallel=L-{1\over 2}\int_0^L dt\;\vec{v}(t)^2$, and from $Z$ determine the mean, variance, and distribution of $R_\parallel$.

The properties of a semi-flexible polymer in cylindrical channels, but without the longitudinal stretching force, have been studied by simulations \cite{dfl,bb,wg,cs,ybg,cbb,twf,byg} and calculated theoretically \cite{lm,twf,byg,twb95} with the channel replaced by a parabolic confining potential. The dynamical evolution of the confined polymer under sudden changes of a longitudinal stretching force or relaxation of the confinement is studied in Ref. \cite{tof}.

In this paper we assume that the confining potential in the Hamiltonian (\ref{Hamiltonian}) has the parabolic form
\begin{equation}
V(\vec{r})={1\over 2}\left(b_x x^2+b_y y^2\right).\label{parabpot}
\end{equation}
In Section II an exact analytical result is obtained for the corresponding partition function $Z\left(\vec{r},\vec{v};\vec{r}_0,\vec{v}_0;t\right)$ for a polymer of contour length $t=L$, with arbitrary fixed values $\vec{r},\vec{v}$ and $\vec{r}_0,\vec{v}_0$ of the position and slope at $t$ and at $t=0$, respectively. An analogous result, but without the second, longitudinal force term in the Hamiltonian (\ref{Hamiltonian}), was derived in Ref. \cite{twb95}. Here we consider the complete quadratic Hamiltonian, quadratic in all three quantities $d\vec{r}/dt$, $d\vec{r}/dt$ and $\vec{r}$. We note that the boundary condition of fixed end points and end slopes is more general than the boundary conditions considered in Refs. \cite{byg,lm}, which are basically periodic, so that the Hamiltonian (\ref{Hamiltonian}), expressed in terms of Fourier components of $\vec{r}(t)$, is diagonal. In Section III compact analytical expressions for the partition function for those and several other boundary conditions, including the experimentally relevant case of freely fluctuating ends, are derived from our result for $Z\left(\vec{r},\vec{v};\vec{r}_0,\vec{v}_0;t\right)$. In Section IV we study the equilibrium response of the polymer to the longitudinal stretching force and show how to calculate the endpoint distribution or radial distribution function by making the replacement $f\to f+s$ in the partition function and then performing the inverse Laplace transformation $s\to L-R_\parallel$. Section V contains closing remarks.

\section{Evaluation of the partition function}
For the Hamiltonian (\ref{Hamiltonian}) with confining potential (\ref{parabpot}), the fluctuations in the $x$ and $y$ directions are statistically independent, and the partition function factors in the form
\begin{equation}
Z\left(\vec{r},\vec{v};\vec{r}_0,\vec{v}_0;t\right)=Z^{(2)}_{f,b_x}(x,v_x;x_0,v_{x0};t)
Z^{(2)}_{f,b_y}(y,v_y;y_0,v_{y0};t),\label{Z(3)}
\end{equation}
where
\begin{equation}
Z^{(2)}_{f,b}\left(x,v;x_0,v_0;t\right)=\int Dx\exp\left\{-{1\over 2}\int_0^t dt\left[P\left({d^2 x\over dt^2}\right)^2+f\left({dx\over dt}\right)^2+b x^2\right]\right\},\label{Z(2)}
\end{equation}
is the corresponding path integral in the two-dimensional space $(x,t)$. Except for the term with coefficient $f$ in the action, the path integral (\ref{Z(2)}) is the same as considered in Ref. \cite{twb95} and can be evaluated in the same way. As in the Feynman-Hibbs treatment of the quantum harmonic oscillator \cite{fh}, we substitute $x(t)=x^*(t)+\xi(t)$ in Eq. (\ref{Z(2)}), where of all the paths with the prescribed end points and end slopes, $x^*(t)$ is the ``classical" path which minimizes the action. Since the action is quadratic in $x$ and its first two derivatives and minimized by $x^*(t)$, and since $\xi(t)$ and $d\xi(t)/dt$ vanish at the end points,
\begin{equation}
Z^{(2)}_{f,b}\left(x,v;x_0,v_0;t\right)=Z^{(2)}_{f,b}(0,0;0,0;t)\exp[-S^*(x,v;x_0,v_0,t)],\label{pathintegral1}
\end{equation}
where $S^*(x,v;x_0,v_0,t)$ is the action for the classical path.

The classical path satisfies the Euler-Lagrange equation
\begin{equation}
P{d^4 x^*\over dt^4}-f{d^2x^*\over dt^2}+b x^*=0,\label{classical eqmo}
\end{equation}
which has the general solution
\begin{equation}
x^*(t)=\sum_{k=1}^4 c_ig_i(t)\label{gensol},
\end{equation}
where
\begin{eqnarray}
&&g_1(t)=\cosh\left(\alpha\hat{t}\thinspace\right)\cos\left(\beta\hat{t}\thinspace\right)\label{g1},\\
&&g_2(t)=\cosh\left(\alpha\hat{t}\thinspace\right)\sin\left(\beta\hat{t}\thinspace\right)\label{g2},\\
&&g_3(t)=\sinh\left(\alpha\hat{t}\thinspace\right)\cos\left(\beta\hat{t}\thinspace\right)\label{g3},\\
&&g_4(t)=\sinh\left(\alpha\hat{t}\thinspace\right)\sin\left(\beta\hat{t}\thinspace\right)\label{g1},
\end{eqnarray}
and
\begin{equation}
\alpha=\sqrt{{1+\gamma}\over 2},\quad\beta=\sqrt{{1-\gamma}\over 2},\quad\gamma=\textstyle{{1\over 2}}\thinspace f(bP)^{-1/2},\quad \hat{t}=b^{1/4}P^{-1/4}t.\label{alphabeta}
\end{equation}
Choosing the expansion coefficients $c_1,...,c_4$ in Eq. (\ref{gensol}) so that $x^*(t)$ reproduces the prescribed positions and slopes at the ends of the polymer and then evaluating the action, we obtain
\begin{eqnarray}
&&S^*(x,v;x_0,v_0,t)=D^{-1}\qquad\qquad\qquad\qquad\nonumber\\
&&\quad\times\left\{\left[\alpha^{-1}\sinh\left(2\alpha\hat{t}\thinspace\right)+
\beta^{-1}\sin\left(2\beta\hat{t}\thinspace\right)
\right]\left(\hat{x}^2+\hat{x}_0^2\right)\right.\nonumber\\
&&\qquad -2\left[\alpha^{-2}\sinh^2\left(\alpha\hat{t}\thinspace\right)
+\beta^{-2}\sin^2\left(\beta\hat{t}\thinspace\right)\right]\left(xv-x_0v_0\right)\nonumber\\
&&\qquad +\left[\alpha^{-1}\sinh\left(2\alpha\hat{t}\thinspace\right)-
\beta^{-1}\sin\left(2\beta\hat{t}\thinspace\right)
\right]\left(\hat{v}^2+\hat{v}_0^2\right)\nonumber\\
&&\qquad -4\left[\beta^{-1}\cosh\left(\alpha\hat{t}\thinspace\right)\sin\left(\beta\hat{t}\thinspace\right)
+\alpha^{-1}\sinh\left(\alpha\hat{t}\thinspace\right)\cos\left(\beta\hat{t}\thinspace\right)\right]\hat{x}\hat{x}_0
\nonumber\\
&&\qquad -4\left[\alpha^{-1}\beta^{-1}\sinh\left(\alpha\hat{t}\thinspace\right)
\sin\left(\beta\hat{t}\thinspace\right)\right]\left(\hat{x}\hat{v}_0-\hat{v}\hat{x}_0\right)\nonumber\\
&&\qquad\left.+4\left[\beta^{-1}\cosh\left(\alpha\hat{t}\thinspace\right)\sin\left(\beta\hat{t}\thinspace\right)
-\alpha^{-1}\sinh\left(\alpha \hat{t}\thinspace\right)\cos\left(\beta\hat{t}\thinspace\right)\right]\hat{v}\hat{v}_0\right\},\label{Sstar}
\end{eqnarray}
where
\begin{equation}
D=2\left[\alpha^{-2}\sinh^2\left(\alpha\hat{t}\thinspace\right)
-\beta^{-2}\sin^2\left(\beta\hat{t}\thinspace\right)\right],\label{D}
\end{equation}
and
\begin{equation}
\hat{x}=b^{3/8}P^{1/8}x,\quad\hat{v}=b^{1/8}P^{3/8}v.\label{xandvhat}
\end{equation}

The prefactor $Z^{(2)}_{f,b}\left(0,0;0,0;t\right)$ in Eq. (\ref{pathintegral1}) is determined by substituting Eq. (\ref{pathintegral1}), with $S^*$ given by Eqs. (\ref{Sstar}) and (\ref{D}), in the Fokker-Planck equation
\begin{equation} \left({\partial\over\partial
t}+v{\partial\over\partial
x}-{1\over 2P}\thinspace{\partial^2\over
\partial v^2}+{1\over 2}\thinspace b x^2+{1\over 2}\thinspace f v^2\right) Z^{(2)}_{f,b}\left(x,v;x_0,v_0;t\right)=0.\label{fp}
\end{equation}
Solving the resulting differential equation for $Z^{(2)}_{f,b}\left(0,0;0,0;t\right)$ with the initial condition
\begin{equation}
Z^{(2)}_{f,b}\left(x,v;x_0,v_0;0\right)=\delta(x-x_0)\delta(v-v_0),\label{initialcondition}
\end{equation}
we obtain
\begin{eqnarray} &&Z^{(2)}_{f,b}\left(0,0;0,0;t\right)=\pi^{-1}(bP)^{1/2}\left[\alpha^{-2}\sinh^2\left(\alpha\hat{t}\thinspace\right)
-\beta^{-2}\sin^2\left(\beta\hat{t}\thinspace\right)\right]^{-1/2}.\label{Z0000}
\end{eqnarray}

This completes the exact evaluation of the polymer partition function (\ref{Z(3)}). The result is given in Eqs. (\ref{pathintegral1}), (\ref{alphabeta})-(\ref{xandvhat}), and (\ref{Z0000}). We have checked with {\em Mathematica} that it does indeed satisfy the Fokker-Planck equation (\ref{fp}) with initial condition (\ref{initialcondition}). In the limit $f\to 0$ it is consistent with Eq. (13) of Ref. \cite{twb95}, and the value of the unspecified normalization constant ${\cal N}$ in that equation is found to be ${\cal N}=\left(2\pi^2\right)^{-1}$.

The function $\beta$ of the variable $\gamma$ defined in Eq. (\ref{alphabeta}) has a square root branch point at $\gamma={1\over 2}f(bP)^{-1/2}=1$. No singularity in the partition function is expected at this value of $\gamma$, and that there is none is clear from Eqs. (\ref{Sstar}), (\ref{D}), and (\ref{Z0000}). The quantity $\beta$ only occurs in the combinations like $\beta^{-1}\sin\left(\beta\hat{t}\thinspace\right)$ and $\cos\left(\beta\hat{t}\thinspace\right)$. Since only even powers appear in the expansions of these combinations in powers of $\beta$, they are analytic functions of $\gamma$,

\section{Other boundary conditions}
In the remaining portion of the paper we consider five different boundary condition at the ends of the polymer. In the case of two fixed ends with $x=x_0=0$ and $v=v_0=0$, the appropriate partition function is
\begin{equation}
 Z^{(2)}_{f,b}(t)_{\rm fixed,fixed}=Z_{f,b}^{(2)}\left(0,0;0,0;t\right)=
 \pi^{-1}(bP)^{1/2}\left[\alpha^{-2}\sinh^2\left(\alpha\hat{t}\thinspace\right)-\beta^{-2}\sin^2\left(\beta \hat{t}\thinspace\right)\right]^{-1/2},\label{Zfixedfixed}
 \end{equation}
as follows from Eq. (\ref{Z0000}).
For one free and one similarly fixed end, the partition function is
\begin{eqnarray}
 Z^{(2)}_{f,b}(t)_{\rm free,fixed}&=&\int_{-\infty}^\infty dx\int_{-\infty}^\infty dv\thinspace Z_{f,b}^{(2)}\left(x,v;0,0;t\right)\nonumber\\ &=&2
 \left[(1+2\gamma)\alpha^{-2}\sinh^2\left(\alpha\hat{t}\thinspace\right)-(1-2\gamma)\beta^{-2}\sin^2\left(\beta \hat{t}\thinspace\right)+4\right]^{-1/2}.\label{Zfreefixed}
 \end{eqnarray}
In the case of two free ends, most relevant to experiments,
\begin{eqnarray}
 Z^{(2)}_{f,b}(t)_{\rm free,free}&=&\int_{-\infty}^\infty dx\int_{-\infty}^\infty dv\int_{-\infty}^\infty dx_0\int_{-\infty}^\infty dv_0\thinspace Z_{f,b}^{(2)}\left(x,v;x_0,v_0;t\right)\nonumber\\ &=&4\pi(bP)^{-1/2}
 \left[(1+2\gamma)^2\alpha^{-2}\sinh^2\left(\alpha\hat{t}\thinspace\right)-(1-2\gamma)^2\beta^{-2}
 \sin^2\left(\beta \hat{t}\thinspace\right)\right]^{-1/2}.\label{Zfreefree}
 \end{eqnarray}
For periodic boundary conditions
\begin{eqnarray}
Z^{(2)}_{f,b}(t)_{\rm periodic}&=&\int_{-\infty}^\infty dx\int_{-\infty}^\infty dv \int_{-\infty}^\infty\thinspace Z_{f,b}^{(2)}\left(x,v;x,v;t\right)\nonumber\\ &=&{1\over 2}\thinspace
\left[\cosh\left(\alpha\hat{t}\thinspace\right)
-\cos\left(\beta\hat{t}\thinspace\right)\right]^{-1}.\label{Zperiodic}
\end{eqnarray}

In calculating the radial distribution function, Levi and Mecke \cite{lm} allow the polymer end points to fluctuate but with the end slopes fixed to the value zero. The corresponding partition function is given by
\begin{eqnarray}
Z^{(2)}_{f,b}(t)_{\rm lm}&=&\int_{-\infty}^\infty dx\int_{-\infty}^\infty dx_0 \int_{-\infty}^\infty\thinspace Z_{f,b}^{(2)}\left(x,0;x_0,0;t\right)\nonumber\\ &=&b^{-1/4}P^{1/4}\left[\cosh^2\left(\alpha\hat{t}\thinspace\right)
-\cos^2\left(\beta\hat{t}\thinspace\right)\right]^{-1/2}\label{Zlm}
\end{eqnarray}
We have checked that Eq. (\ref{Zlm}) is in complete agreement with the result \cite{lm}
\begin{equation}
{Z^{(2)}_{s+f,b}(t)_{\rm lm}\over Z^{(2)}_{f,b}(t)_{\rm lm}}=\prod_{n=1}^\infty\left[{P(n\pi/t)^4+f(n\pi/t)^2+b\over P(n\pi/t)^4+(s+f)(n\pi/t)^2+b}\right]^{1/2},\label{Zlmlm}
\end{equation}
obtained by substituting $x(t')=\sum_{n=0}^\infty x_n\cos(\pi n t'/t)$, $0<t'<t$ in the Hamiltonian and then performing Gaussian integrals over the coefficients $x_n$ to obtain the partition function.

For periodic boundaries the appropriate Fourier expansion is $x(t')=\sum_{n=-\infty}^\infty x_n e^{2\pi i n t'/t}$. This leads to
\begin{equation}
{Z^{(2)}_{s+f,b}(t)_{\rm per}\over Z^{(2)}_{f,b}(t)_{\rm per}}=\prod_{n=1}^\infty{P(2n\pi/t)^4+f(2n\pi/t)^2+b\over P(2n\pi/t)^4+(s+f)(2n\pi/t)^2+b},\label{Zperper}
\end{equation}
which is completely consistent with our result (\ref{Zperiodic}).

For all of the boundary conditions considered above, $Z^{(2)}_{f,b}$ has the asymptotic behavior
\begin{eqnarray}
&&-\ln Z^{(2)}_{f,b}(t)\approx \alpha\hat{t}=E_0(f,b) t,\label{Zasymptotic1}\\
&&E_0(f,b)=\alpha \left({b\over P}\right)^{1/4} =2^{-1/2}\left({b\over P}\right)^{1/4}\left[1+\textstyle{1\over 2}\thinspace f(bP)^{-1/2}\right]^{1/2},\label{Zasymptotic2}
\end{eqnarray}
both in the long-polymer regime $t\to\infty$ with $f$ fixed and in the strong-stretching regime $f\to\infty$ with $t$ finite. In the latter regime $E_0(f,b)$ in Eq. (\ref{Zasymptotic2}) further simplifies to $E_0(f,b)\approx {1\over 2}\thinspace (f/P)^{1/2}$. As discussed in the appendix of Ref. \cite{byg}, the dominant contribution to $Z^{(2)}_{f,b}$ for large $t$ comes from the ``ground state" of the $t$-independent Fokker-Planck equation, which has eigenvalue $E_0(f,b)$.

\section{Moments and radial distribution function}
According to Eqs. (\ref{Z(3)}) and (\ref{Z(2)}), the mean value and variance of the difference $\xi=L-R_\parallel
={1\over 2}\int_0^t dt\;\vec{v}(t)^2$ of the contour and longitudinal lengths are given by
\begin{eqnarray}
&&\langle\xi\rangle=-{\partial\over\partial f}\ln\left[Z^{(2)}_{f,b_x}(t)Z^{(2)}_{f,b_y}(t)\right]\,\label{avxi}\\
&&\langle\left(\xi-\langle\xi\rangle\right)^2\rangle={\partial^2\over\partial f^2}\ln\left[Z^{(2)}_{f,b_x}(t)Z^{(2)}_{f,b_y}(t)\right]\label{varxi}
\end{eqnarray}
In Figs. 1 and 2, these two moments are plotted as functions of the contour length $t=L$ and the longitudinal force parameter $f=\tau/k_BT$ for all five boundary conditions considered above. In both figures $b_x=b_y=262P^{-3}$. As explained in footnote \cite{explain262}, these are appropriate potential parameters for a long polymer confined in a channel with a square $D\times D$ cross section with $D={1\over 3}P$.

That the lowermost and uppermost curves in Figs. 1 and 2 correspond to fixed-fixed and free-free boundary conditions at the ends of the polymer is reasonable, since these boundary conditions most and least restrict the polymer fluctuations, respectively.

The leading asymptotic forms for $\langle\xi\rangle$ and $\langle\left(\xi-\langle\xi\rangle\right)^2\rangle$ for large $t$ and/or large $f$, toward which the curves in Fig. 1 and 2 tend, are the same for all five boundary conditions and given by
\begin{eqnarray}
&&\langle\xi\rangle\approx 2^{-5/2}P^{-3/4}\left\{b_x^{-1/4}\left[1+\textstyle{1\over 2}\thinspace f(b_xP)^{-1/2}\right]^{-1/2}+\left(b_x\to b_y\right)\right\}t,\label{avasymptotic}\\
&&\langle\left(\xi-\langle\xi\rangle\right)^2\rangle\approx 2^{-9/2}P^{-5/4}\left\{b_x^{-3/4}\left[1+
\textstyle{1\over 2}\thinspace f(b_xP)^{-1/2}\right]^{-3/2}+\left(b_x\to b_y\right)\right\}t,\label{varasymptotic}
\end{eqnarray}
as follows from Eqs. (\ref{Zasymptotic1})-(\ref{varxi}). In the strong-stretching regime $f\to\infty$,
Eqs. (\ref{avasymptotic}) and (\ref{varasymptotic}) further simplify to
\begin{equation}
\langle\xi\rangle\approx\textstyle{1\over 2}\thinspace (fP)^{-1/2}t,\quad
\langle\left(\xi-\langle\xi\rangle\right)^2\rangle\approx {1\over 4}\thinspace f^{-3/2}P^{-1/2}t,\label{avvarlargea}
\end{equation}
independent of the potential parameters $b_x$ and $b_y$, as expected, and in agreement with the result of Marco and Siggia \cite{ms} for a strongly stretched, unconfined semiflexible polymer.

From Eqs. (\ref{extension}) and (\ref{avasymptotic}) we see that
\begin{equation}
\overline{v^2}\equiv{1\over t}\int_0^t dt'\thinspace\vec{v}(t')^2\approx
\left(8b_x^{1/2}P^{3/2}+4fP\right)^{-1/2}+\left(b_x\to b_y\right)\label{avofvsq}
\end{equation}
for large $t$. From this relation one can identify the range of parameters $b_x$, $b_y$, $P$, and $f$ consistent with the domain of validity $\overline{v^2}\ll 1$ of our results. For $b_x=b_y=262P^{-3}$ and $f=0$, as considered in Figs. 1-3, Eq. (\ref{avofvsq}) yields
$\overline{v^2}=0.176$.

We now turn to the probability distribution or radial distribution function
\begin{equation}
P_f(\xi)=\left\langle\delta\left(\xi-{1\over 2}\int_0^Ldt\thinspace \vec{v}(t)^2\right)\right\rangle\;,\label{P(z)}
\end{equation}
from which the above moments follow. Its Laplace transform is given by
\begin{equation}
\tilde{P}_f(s)=\int_0^\infty d\xi\thinspace e^{-s\xi}P_f(\xi)=\left\langle\exp\left(-{s\over 2}
\int_0^Ldt\thinspace\vec{v}(t)^2\right)\right\rangle\;.\label{Ptilde1}
\end{equation}
Comparing this expression with Eqs. (\ref{Z(3)}) and (\ref{Z(2)}), we see that
\begin{equation}
\tilde{P}_f(s)={Z_{f+s,b_x}^{(2)}(t)\over Z_{f,b_x}^{(2)}(t)}{Z_{f+s,b_y}^{(2)}(t)\over Z_{f,b_y}^{(2)}(t)}.
\label{Ptilde2}
\end{equation}

Substituting the partition functions (\ref{Zfixedfixed})-(\ref{Zlm}) in Eq. (\ref{Ptilde2}), we obtain  simple analytic expressions for $\tilde{P}_f(s)$ for each of the five boundary conditions considered above. The various moments of $\xi$ may be generated from $\tilde{P}_f(s)$ according to
\begin{equation}
\langle\xi^n\rangle=\left.\left(-\partial/\partial s\right)^n\tilde{P}_f(s)\right\vert_{s=0},\quad
\langle\left(\xi-\langle\xi\rangle\right)^n\rangle=\left.\left(-\partial/\partial s\right)^n\ln \tilde{P}_f(s)\right\vert_{s=0},
\end{equation}
as follows from Eq. (\ref{Ptilde1}),
and the radial distribution function $P_f(\xi)$ is determined by the inverse Laplace transformation
\begin{equation}
P_f(\xi)={1\over 2\pi i}\int_{c-i\infty}^{c+i\infty}ds\thinspace{\tilde P}_f(s)e^{\xi s}.\label{inversetransform}
\end{equation}

Since $\tilde{P}_f(s)$, as given by Eq. (\ref{Ptilde2}), only depends on $s$ in the combination $s+f$, Eq. (\ref{inversetransform}) implies
\begin{equation}
P_f(\xi)=A_fe^{-f\xi}P_0(\xi).\label{PfP0}
\end{equation}
Thus, the radial distributions in the presence and absence of the stretching force only differ by a factor
$e^{-f\xi}$ and a corresponding normalization constant $A_f$. From Eq. (\ref{Ptilde1}) and the normalization $\int_0^\infty d\xi\thinspace P_f(\xi)=1$, we see that $A_f={\tilde P}_0(f)^{-1}$, in terms of the quantity $\tilde{P}_f(s)$ defined in Eq. (\ref{Ptilde2}).

For the boundary condition that the polymer slope vanish at the end points and for $f=0$ and $b_x=b_y=b$, Levi and Mecke \cite{lm} obtained an analytical expression in the form of a power series  for the radial distribution function $P_f(\xi)$ by substituting Eq. (\ref{Zlmlm}) in Eq. (\ref{Ptilde2}) and performing the integration over $s$ in Eq. (\ref{inversetransform}) using Cauchy's residue theorem. For the periodic boundary condition $P_f(\xi)$ can be evaluated analytically, beginning with Eq. (\ref{Zperper}), in the same way.

For all five boundary conditions, for $L=P$ and $f=0$, and for the same potential parameters $b_x=b_y=262P^{-3}$ as in Figs. 1 and 2, we have  evaluated $P_f(\xi)$ by performing the inverse Laplace transform (\ref{inversetransform}) numerically using Stehfest's method \cite{hs}. This leads to then distributions of $R_\parallel/L$ shown in Fig. 3. As the polymer is fairly short and not so tightly confined, the boundary conditions significantly affect the results. As expected, the narrowest, rightmost peak and the broadest, leftmost peak correspond to fixed-fixed and free-free boundary conditions, respectively.

\section{Concluding remarks}
The properties of a semiflexible polymer confined along a line by a parabolic potential have already been studied in several other papers \cite{ybg,twf,byg,lm,twb95,tof}, and we conclude by summarizing what this paper adds. As in Ref. \cite{twb95}, we concentrate on the regime in which the polymer is stretched out along the line due to a strong confining potential and/or a strong longitudinal force applied to the ends. The main new result is an exact analytical expression for the partition function or path integral $Z\left(\vec{r},\vec{v};\vec{r}_0,\vec{v}_0;t\right)$ , defined by Eqs. (\ref{Z(3)}) and (\ref{Z(2)}),
where the Hamiltonian (\ref{Hamiltonian}) is quadratic in all three quantities $d\vec{r}/dt$, $d\vec{r}/dt$ and $\vec{r}$. Inclusion of the term quadratic in $d\vec{r}/dt$, with coefficient $f$, allows us to study the response of the polymer to a longitudinal stretching force and to calculate the distribution of the end-to-end distance or radial distribution function by replacing $f$ by $f+s$ in the partition function and performing the inverse Laplace transformation $s\to\xi=L-R_\parallel$, as in Eqs. (\ref{Ptilde2}) and (\ref{inversetransform}).

From our result for the partition function $Z\left(\vec{r},\vec{v};\vec{r}_0,\vec{v}_0;t\right)$ for arbitrary fixed endpoints and endslopes, we obtain compact analytical expressions (\ref{Zfixedfixed})-(\ref{Zlm}) for the partition functions corresponding to five other boundary conditions of interest, including the case of freely fluctuating ends most relevant to experiments. From these expressions it is a simple to calculate the radial distribution function by numerical inversion of the Laplace transform. In Fig. 1-3 we  present numerical results, obtained this way, for the radial distribution function and its first two moments, for polymers which are short enough ($L=P=3D)$, so that the effects of the different boundary conditions at the ends are appreciable.

\newpage
\begin{figure}[Figure1]
\begin{center}
\includegraphics[width=5.0in]{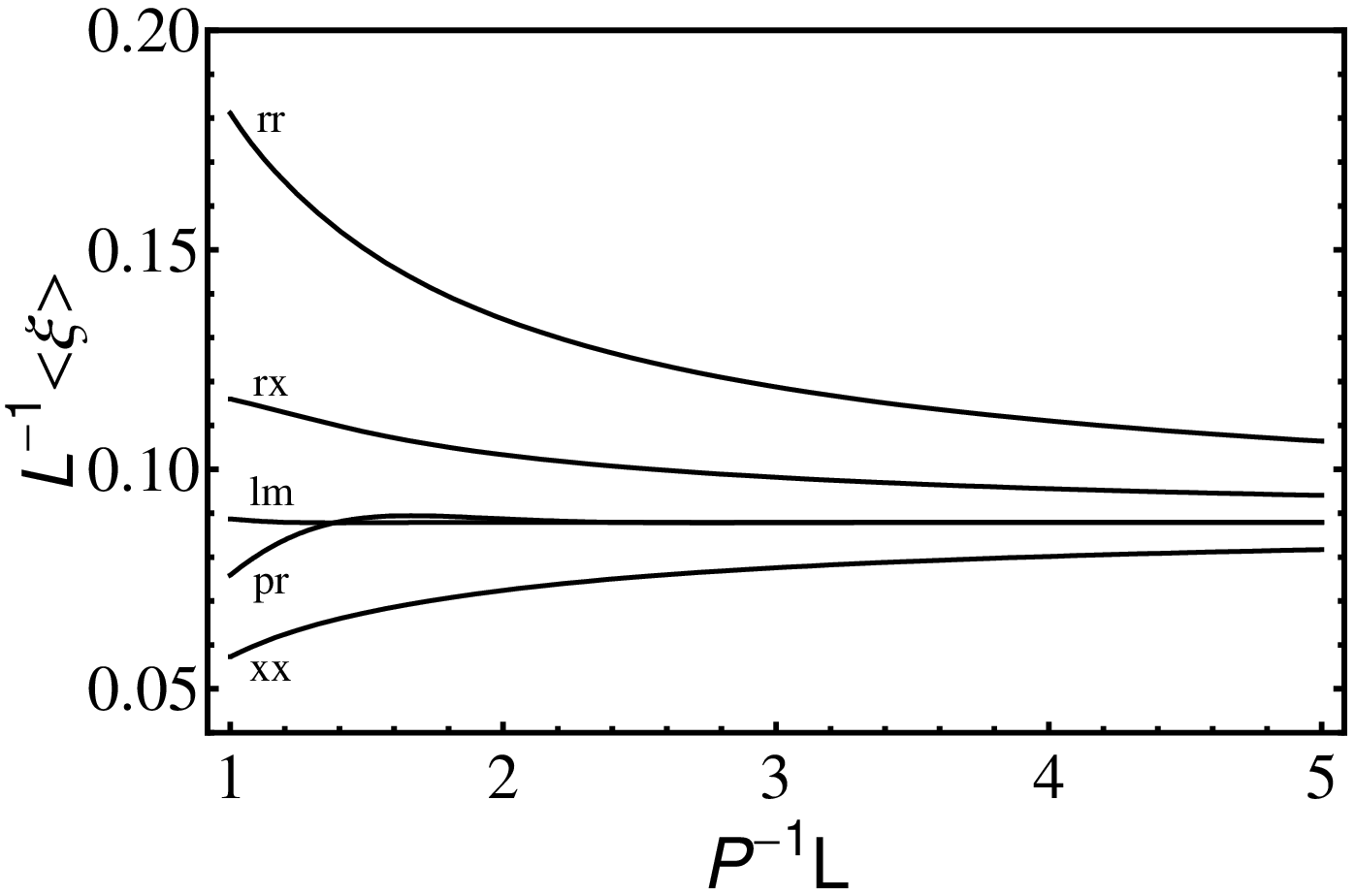}\vskip 1.2 cm
\includegraphics[width=5.0in]{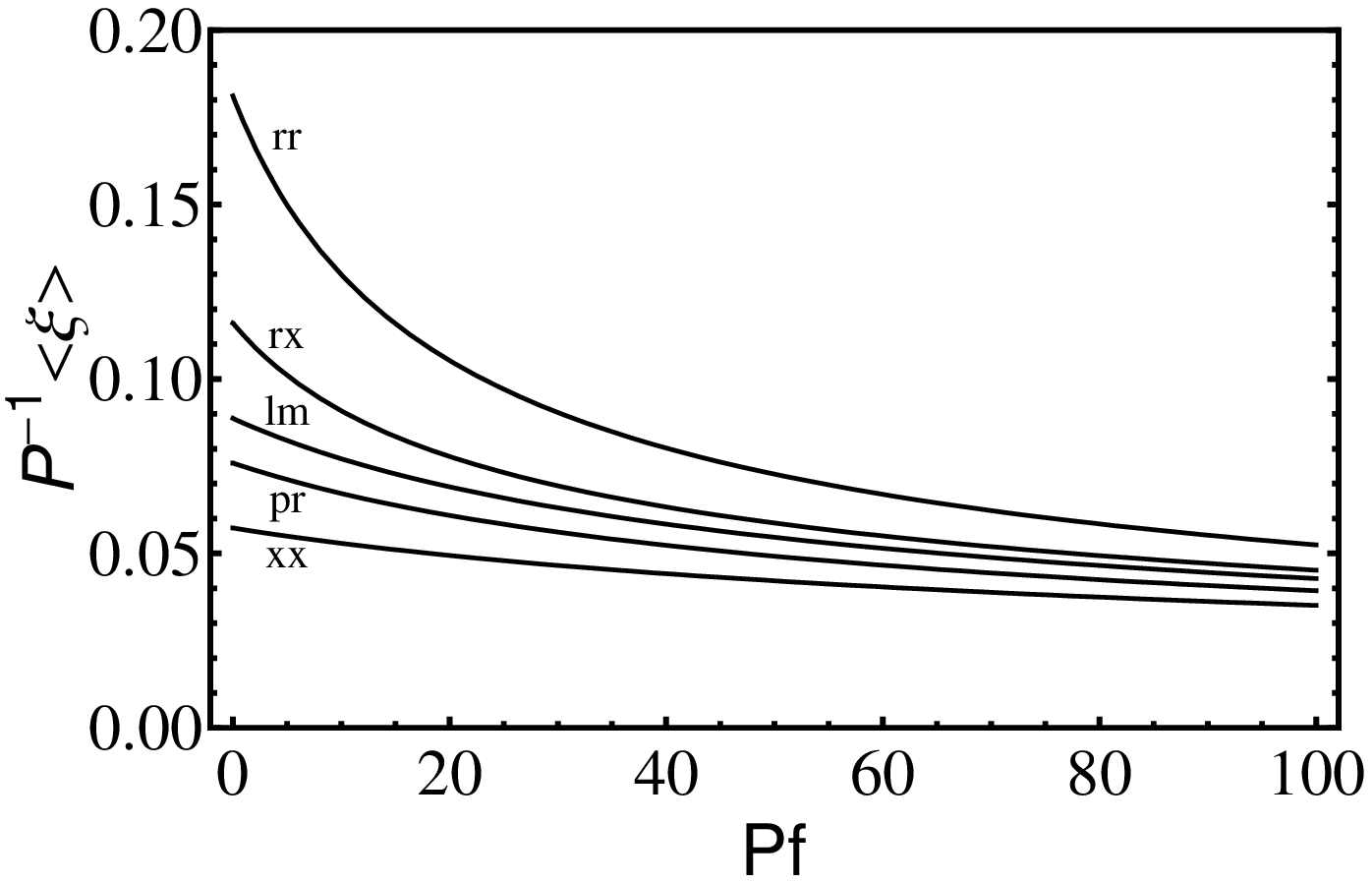}
\caption{Dependence of $\langle\xi\rangle=L-\langle R_\parallel\rangle$, where $L$ is the contour length and $R_\parallel$ is the longitudinal length of the polymer, for fixed persistence length $P$, on $L$ (upper figure) and on the longitudinal force parameter $f=\tau/k_BT$ (lower figure). As explained following Eq. (\ref{varxi}), the potential parameters $b_x=b_y=262P^{-3}$ used for the figures are appropriate for a polymer in a channel with a square $D\times D$ cross section with $D={1\over 3}P$. In the upper figure $f=0$, and in the lower figure $L=P$. The curves are labeled according to the boundary condition at the ends of the polymer: free-free (rr), free-fixed (rx), Levi-Mecke (lm), periodic (pr), fixed-fixed (xx).}\label{fig1}
\end{center}
\end{figure}

\newpage
\begin{figure}[Figure2]
\begin{center}
\includegraphics[width=5.0in]{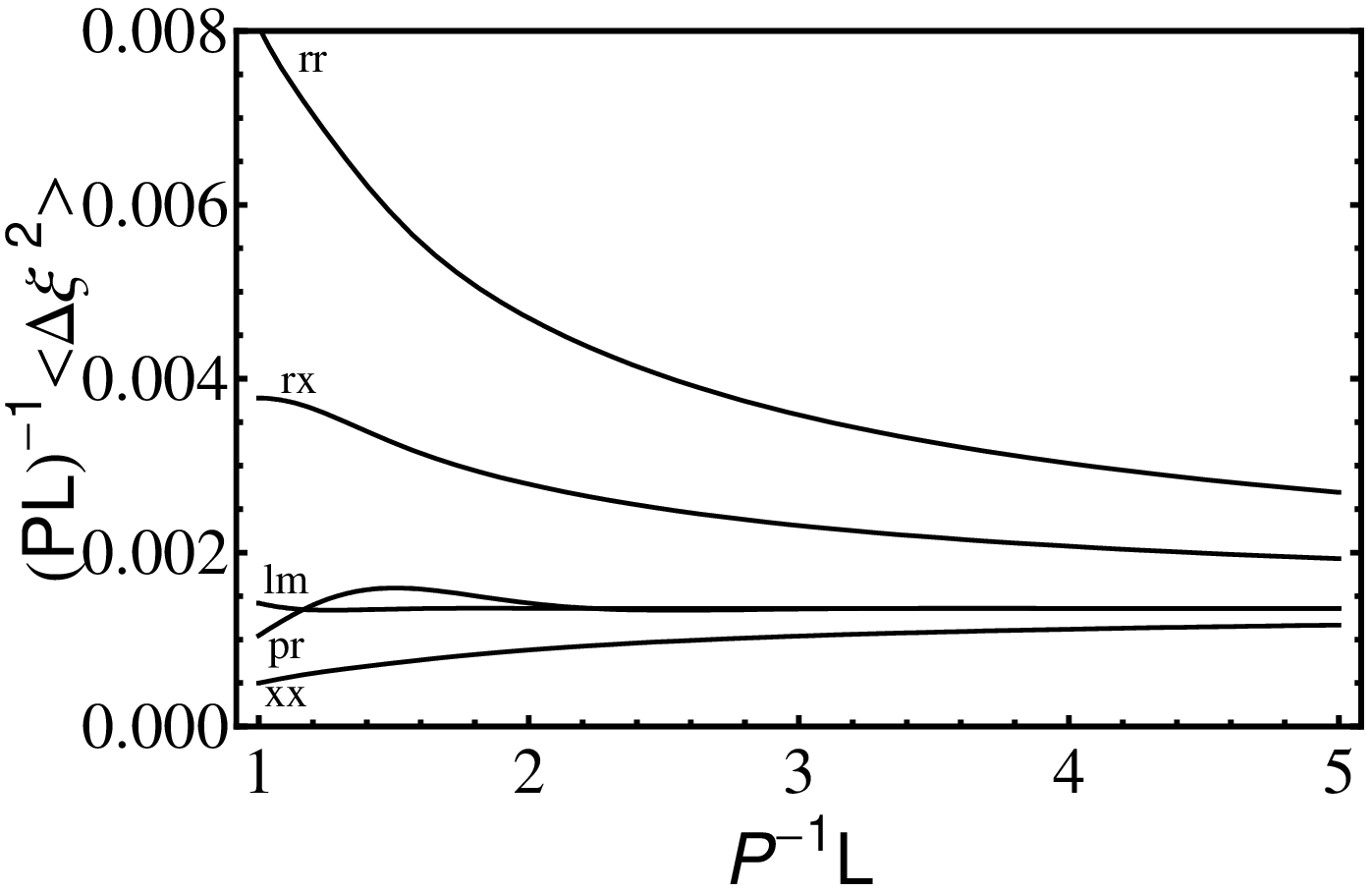}\vskip 1.2 cm
\includegraphics[width=5.0in]{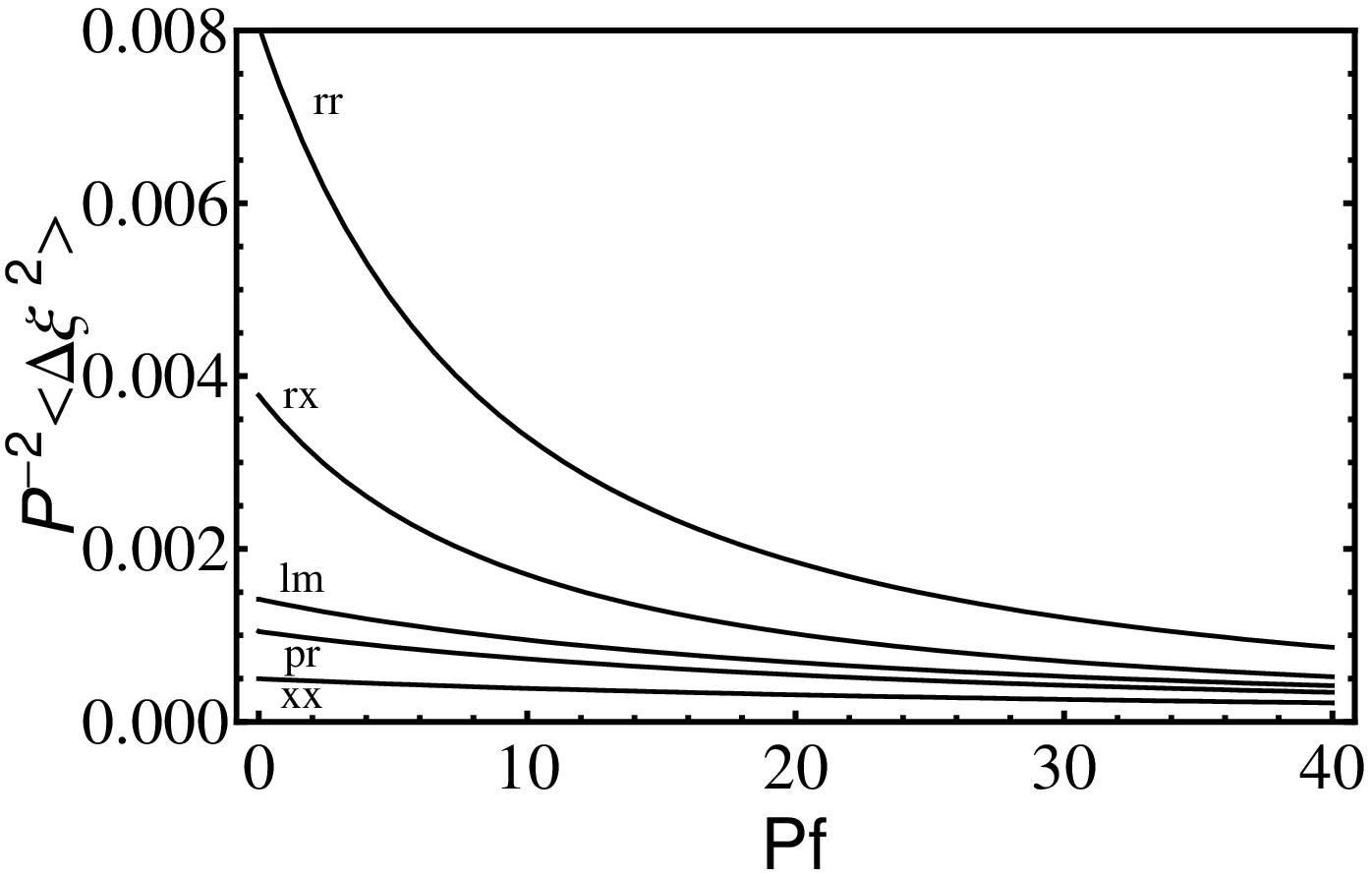}
\caption{Same as Fig. 1 except that the variance $\langle\Delta\xi^2\rangle=\langle\left(\xi-\langle\xi\rangle\right)^2\rangle$ is shown.}\label{fig2}
\end{center}
\end{figure}

\newpage
\begin{figure}[Figure3]
\begin{center}
\includegraphics[width=6.0in]{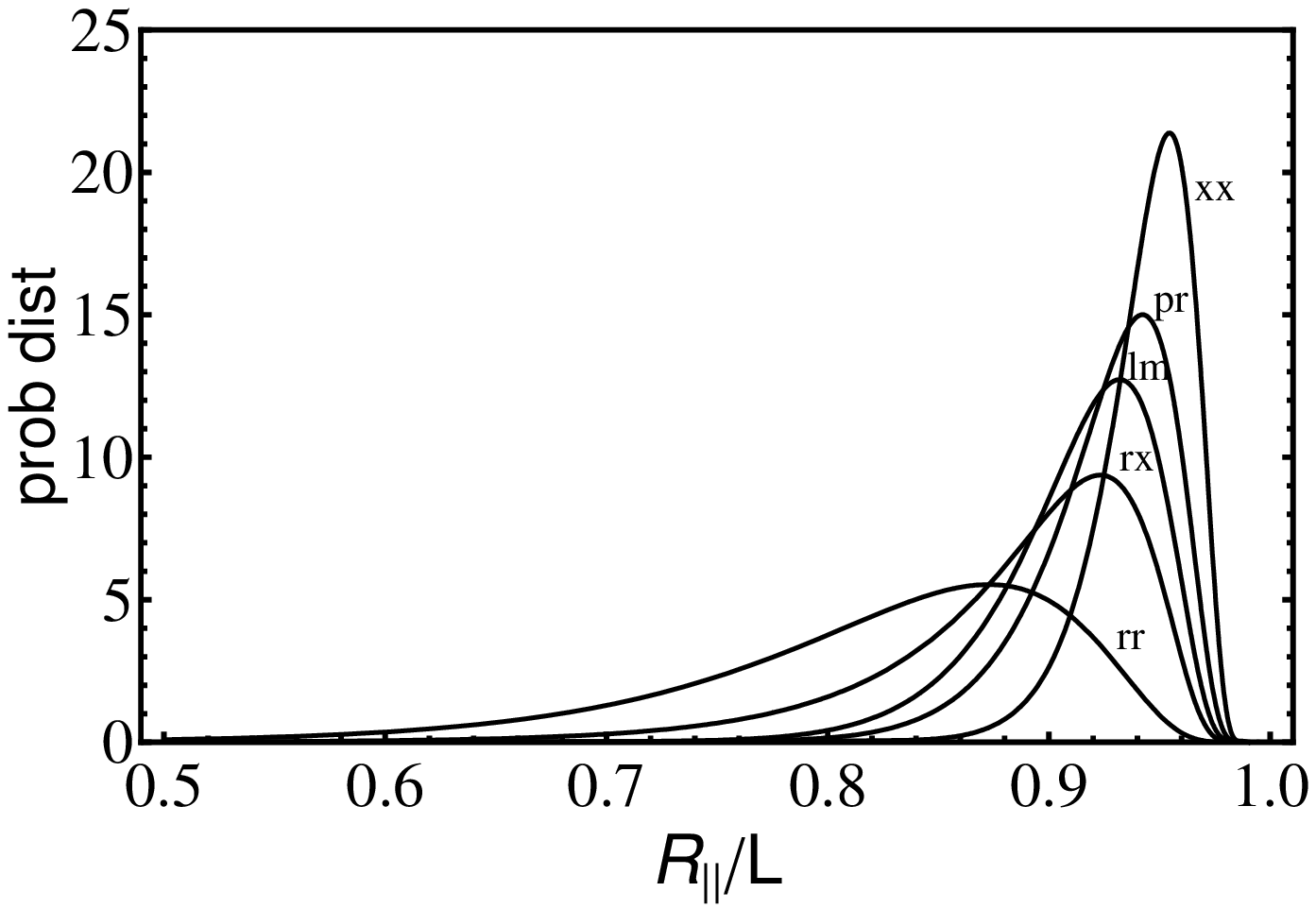}

\caption{Distribution of $R_\parallel/L$ for a polymer with contour length $L=P$ with longitudinal force parameter $f=0$. As in Figs. 1 and 2, the potential parameters are $b_x=b_y=262P^{-3}$, appropriate for a long polymer in a channel with a square $D\times D$ cross section with $D={1\over 3}P$. The curves are labeled according to the boundary condition at the ends of the polymer: free-free (rr), free-fixed (rx), Levi-Mecke (lm), periodic (pr), fixed-fixed (xx).}\label{fig3}
\end{center}
\end{figure}

\begin{thebibliography}{99}
\bibitem{retal} W. Reisner, K. J. Morton, R. Riehn, Y. M. Wang, Z. Yu, M. Rosen,
J. C. Sturm, S. Y. Chou, E. Frey, and R. H. Austin, Phys. Rev. Lett. {\bf 94},
196101 (2005).
\bibitem{ksp} S. K\"oster, D. Steinhauser, and T. Pfohl, J. Phys. Condens.
Matter {\bf 17}, S4091 (2005).
\bibitem{kkp} S. K\"oster, J. Kierfeld, and T. Pfohl, Eur. Phys. J. E {\bf 25}, 439 (2008).
\bibitem{kp} S. K\"oster and T. Pfohl, Cell Motility and the Cytoskeleton {\bf 66}, 771 (2009).
\bibitem{hlgr} M. B. Hochrein, J. A. Leierseder, L. Golubovic, and
J. O. R\"adler, Phys. Rev. Lett. {\bf 96}, 038103 (2006).
\bibitem{chemsocrevpt} F. Persson and J. O. Tegenfeldt, Chem. Soc. Rev. {\bf 39}, 985 (2010), and references therein.
\bibitem{chemsocrevlc} S. L. Levy and H. G. Craighead, Chem. Soc. Rev. {\bf 39}, 1133 (2010), and references therein.
\bibitem{wr} W. Reisner, N. B. Larsen, A. Silahtaroglu, A. Kristensen, N. Tommerup, J. O. Tegenfeldt, and H. Flyvbjerg, Proc. Natl. Acad. Sci. USA {\bf 107}, 13294 (2010).
\bibitem{explain} The bending energy of a worm-like chain is given by
${\cal H}_{\rm bend}={1\over 2}\kappa\int_0^L ds\thinspace\left(d\hat{\tau}/ds\right)^2$,
where $\hat{\tau}=d\vec{r}/ds$ is a unit tangent vector and $s$ denotes the arc length.
Rewritten in terms of the quantities $\vec{v}=d\vec{r}/dt$ and $\vec{a}=d^2\vec{r}/dt^2$, ${\cal H}_{\rm bend}={1\over 2}\kappa\int_0^L dt
\left[(1+\vec{v}^2)^{-3/2}\thinspace\vec{a}^2-(1+\vec{v}^2)^{-5/2}\thinspace(\vec{v}\cdot\vec{a})^2\right]$. For $|\vec{v}|\ll 1$, this simplifies to the bending energy in Eq. (\ref{Hamiltonian}).
\bibitem{dfl} M. Dijkstra, D. Frenkel, and H. N. W. Lekkerkerker,
Physica A {\bf 193}, 374 (1993).
\bibitem{bb} D. J. Bicout and T. W. Burkhardt, J. Phys. A {\bf 34}, 5745 (2001).
\bibitem{wg} J. Wang and H. Gao, J. Chem. Phys. {\bf 123}, 084906 (2005).
\bibitem{cs} J. Z. Y. Chen and D. E. Sullivan, Macromolecules {\bf 39}, 7769 (2006).
\bibitem{ybg} Y. Yang, T. W. Burkhardt, and G. Gompper, Phys. Rev. E {\bf 76}, 011804 (2007).
\bibitem{cbb} P. Cifra, Z. Benkov\'a, and T. Bleha, J. Phys. Chem. B {\bf 113}, 1843 (2009).
\bibitem{twf} F. Th\"uroff, F. Wagner, and E. Frey, EPL {\bf 91}, 38004 (2010).
\bibitem{byg} T. W. Burkhardt, Y. Yang, and G. Gompper, Phys. Rev. E {\bf 82}, 041801 (2010).
\bibitem{twb95} T. W. Burkhardt, J. Phys. A {\bf 28}, L629 (1995).
\bibitem{lm} P. Levi and K. Mecke, EPL {\bf 78}, 38001 (2007).
\bibitem{tof} F. Th\"uroff, B. Obermayer, and E. Frey, Phys. Rev. E {\bf 83}, 021802 (2011).
\bibitem{fh} {\it R. P. Feynman and A. R. Hibbs, Quantum Mechanics and Path Integrals} (McGraw-Hill, New York, 1965).
\bibitem{explain262} According to simulations \cite{ybg,byg}, for a long, tightly confined
polymer in a cylindrical channel with a square $D\times D$ cross section and with no stretching force, $\langle\xi\rangle\approx 2(0.09137\pm 0.00007)(D/P)^{2/3}t$. For $f=0$ and $b_x=b_y=b$, Eq. (\ref{avasymptotic}) takes the form $\langle\xi\rangle\approx 2^{-3/2}P^{-3/4}b^{-1/4}t$.
Equating these two expressions and choosing $D={1\over 3}P$ leads to the effective potential parameters $b_x=b_y=262P^{-3}$ used in Figs. 1-3.
\bibitem{ms} J. F. Marco and E. D. Siggia, Macromolecules {\bf 28}, 8759 (1995).
\bibitem{hs} H. Stehfest, Comm. ACM {\bf 13}, 47 and 624 (1970).
\end{thebibliography}
\end{document}